\documentclass[twocolumn]{aastex62}

\graphicspath{{./}}
\usepackage{color,soul}
\usepackage{multirow}
\usepackage{amsmath}

\received{April 19, 2019}
\revised{May 10, 2019}
\accepted{May 12, 2019}
\submitjournal{ApJL}

\shorttitle{Deep Classifier with Novelty Detection}
\shortauthors{B. T.-H. Tsang and W. C. Schultz}

\begin{document}

\title{Deep Neural Network Classifier for Variable Stars with Novelty Detection Capability}

\correspondingauthor{Benny T.-H. Tsang}
\email{btsang@kitp.ucsb.edu}

\author{Benny T.-H. Tsang}
\affil{Kavli Institute for Theoretical Physics, University of California, Santa Barbara, CA 93106, USA}

\author{William C. Schultz}
\affiliation{Department of Physics, University of California, Santa Barbara, CA 93106, USA}



\begin{abstract}

Common variable star classifiers are built only with the goal of 
producing the correct class labels,
leaving much of the multi-task capability of deep neural networks unexplored. 
We present a periodic light curve classifier that combines a recurrent neural
network autoencoder for unsupervised feature extraction and 
a dual-purpose estimation network for supervised classification and novelty detection. 
The estimation network optimizes a Gaussian mixture model in the reduced-dimension 
feature space, where each Gaussian component corresponds to a variable class.
An estimation network with a basic structure of a single hidden layer attains 
a cross-validation classification accuracy of $\sim99\%$, 
on par with the conventional workhorses, random forest classifiers.
With the addition of photometric features, the network is capable of detecting
previously unseen types of variability with 
precision 0.90, recall 0.96, and an $F_{1}$ score of 0.93.
The simultaneous training of the autoencoder and estimation network is found to 
be mutually beneficial, resulting in faster autoencoder convergence, and superior classification and novelty detection performance.
The estimation network also delivers adequate results even when optimized with 
pre-trained autoencoder features,
suggesting that it can readily extend existing classifiers to provide 
added novelty detection capabilities.

\end{abstract}


\section{Introduction}
\label{sec:intro}
Efficient classification of the variability of astrophysical objects is crucial to 
defining follow-up observations and analysis. 
With the advent of the next-generation surveys such as the 
Large Synoptic Survey Telescope \citep[LSST,][]{Ivezic08,LSST09}
and the Zwicky Transient Facility \citep[ZTF][]{ZTF19},
automatic pipelines are required to categorize an unprecedented amount of 
light curves into known or previously unseen variability classes.
To this end, machine learning has been applied to solve the
classification problem sufficiently. The identification of objects with novel variability 
properties, however, still relies heavily on visual inspection by human experts.  

Classifiers are seldom trained using raw light curves due to their high 
and non-uniform number of observational epochs. 
Instead, `features' of much lower dimensions are obtained, a process known as
\emph{feature extraction}.
Features typically include the variable period \citep{Lomb76,Scargle82,SC96,BLS02}, 
amplitudes and ratios of different Fourier components \citep{Nun+15}, 
and summary statistics of the flux variation (e.g. standard deviation and skewness).
Random forest (RF) classifiers trained on features obtained from photometric data
have been the blueprint of most recent classification efforts
\citep{Richards+11,Dubath+11,Bloom+12,KPBJ14,Masci+14,KBJ16,Jayasinghe+18b,Rimoldini18}.
The new probabilistic RF method, which takes into account uncertainties in both features and labels 
\citep{RBS19}, holds promise in improving RF's performance.
However, manual selection of features still requires tremendous human involvement. 


Deep artificial neural networks have attracted recent attention in the 
physical sciences due to their ability to acquire meaningful data representations
with minimal human input.
In astrophysics alone, galaxy morphology classification \citep{DWD15,AT17},
transient and exo-planet detection \citep{CV+17,SV18,SM18},
and, of course, light curve classification \citep{APB19,Muthukrishna19}
have benefited from the success of convolutional neural networks (CNNs) in image and sequential data processing. 

Autoencoding recurrent neural networks (RNNs),
which preserve the sequential information of input data, 
were found to be effective in extracting representative features from
light curves \citep{Naul+18}. 
The encoder component of the network takes light curves as inputs and generates 
features of much lower dimension.
The decoder then attempts to reconstruct the light curves using only the encoder-generated features.
By matching the reconstructed light curves with the original inputs,
the autoencoder learns to isolate essential features that characterize the light curves.
Unlike the conventional approach where frequency and statistical features 
are hand-selected, the RNNs perform feature extraction 
without human intervention.

Most of the previous attempts at light curve classification focused only on
correctly providing object labels. 
In a realistic workflow, however, an indispensable task is to uncover objects with previously unseen
types of variability, so called \emph{novelty detection}, 
among a large number of objects of known classes. 
\citet{Zong+18} have recently proposed a framework that combines autoencoding 
feature extraction with a Gaussian Mixture Model (GMM)-based 
unsupervised anomaly detection scheme. 

In this letter, we present a neural network architecture that combines 
the efforts of \citet{Naul+18} and \citet{Zong+18}. 
Namely, we jointly optimize an autoencoding RNN for feature extraction
from variable light curves and an estimation network for 
classification and novelty detection. The motivation for this work is to promote the application of multi-task neural networks in variability analysis.

\begin{figure*}
  \begin{center}
  \includegraphics[trim=3cm 0cm 3cm 3cm, clip, width=\textwidth]{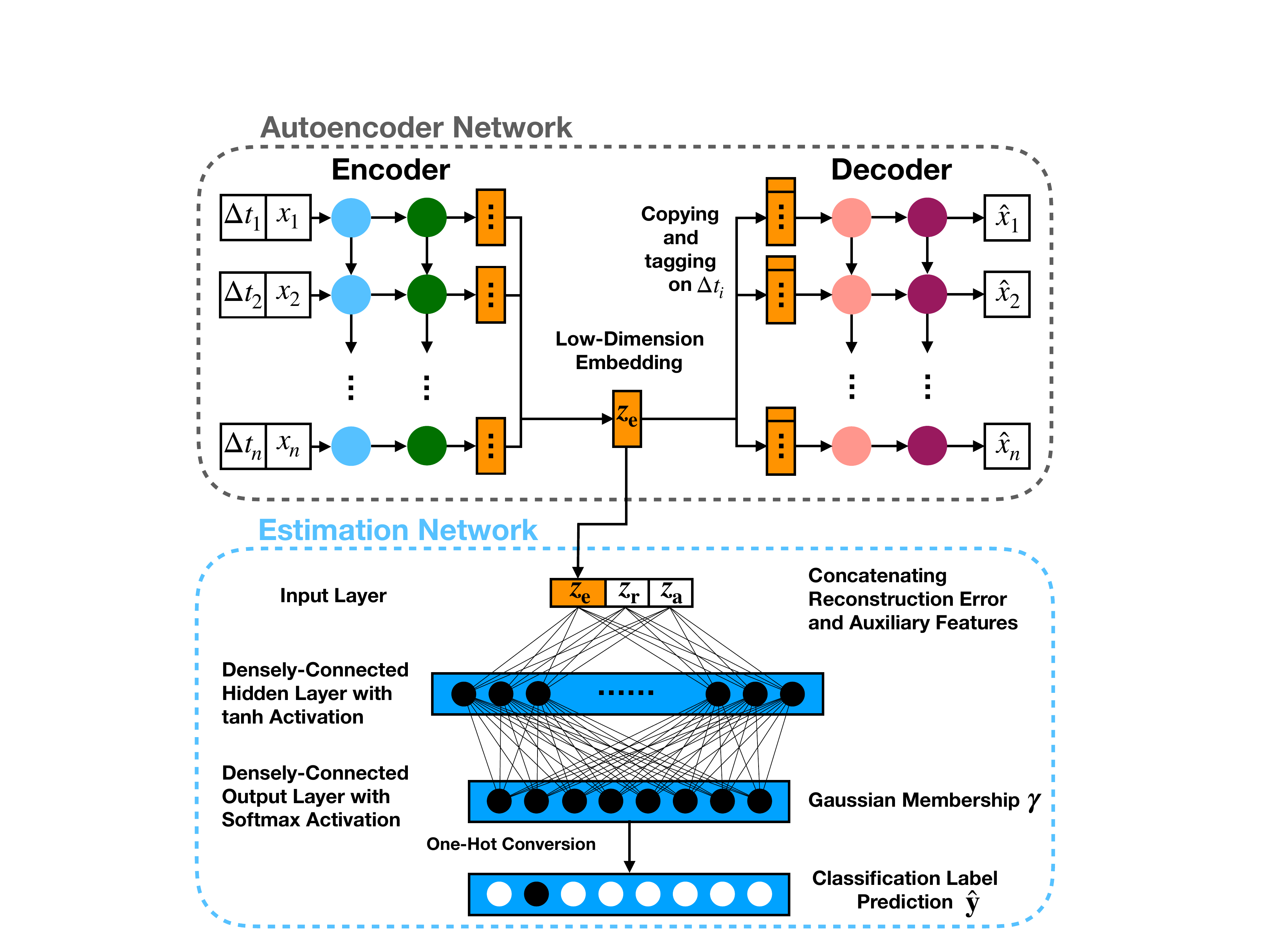}
  \end{center}
  \caption{Schematic diagram of the neural network architecture.
           The structure of the RNN autoencoder follows \citet{Naul+18}.
           The estimation network is similar to that used in \citet{Zong+18} except for
           the added one-hot conversion for the classification task.}
  \label{fig:NN_schematic}
\end{figure*}

\section{Methods}
\label{sec:methods}
\subsection{Data and data pre-processing}
\label{sec:data}
The network is trained using the light curves from the 
All-Sky Automated Survey for Supernovae (ASAS-SN) Variable Stars Database
I and II \citep{Shappee+14,Jayasinghe+18b}. 
The light curves, obtained from the online database\footnote{\href{https://asas-sn.osu.edu/variables}{https://asas-sn.osu.edu/variables}}, typically have hundreds of epochs in V and g bands.
From the vast database, only light curves of types listed in Table \ref{tab:classnames}
were selected for classification purposes.
These variability types will be referred to as the variable \emph{superclasses}.

The data selection and pre-processing procedure closely resembles that of \cite{Naul+18}.
To minimize confusion, 
we selected sources based on information from the \textsc{ASAS-SN} variable star catalog. 
Classifications in the catalog, generated using their RF classifier, are treated as the true labels of the sources.
We note that these labels may not be completely genuine. 
Only variables with classification probabilities above 90\%
and at least 200 epochs were used.
Light curves with \href{https://github.com/jakevdp/supersmoother}{\textsc{SuperSmoother}} residuals greater than 0.7 were omitted to ensure only true periodic sources were included. 
Following \citet{Jayasinghe+18b}, 
we further filtered out saturated and faint sources with $V<11$ and $V>17$. 

In the ASAS-SN variable star database, sources that do not meet any of the classification criteria are referred to as `variable stars of unspecified type' (\textsc{VAR}). 
This inhomogeneous category contains objects with variability types that are principally different from the superclasses. 
Since their light curves are by selection unlike any of the known classes, 
the \textsc{VAR} objects are ideal for assessing the ability of the networks 
in detecting unseen samples.
Similarly, only \textsc{VAR} sources with at least 200 observational epochs were selected.
No classification probability nor supersmoother residual cuts were applied to the light curve sequences in the
\textsc{VAR} class.

Pre-processing began with the partitioning of light curves into sequences 
of equal lengths, $n=200$, as the autoencoder is optimized to process 
input sequences of fixed dimensions.
The above pre-processing resulted in a reduced dataset of $\sim$48,000
light curve sequences in the superclasses and $\sim$5,300 in the \textsc{VAR} class.
Using periods from the ASAS-SN catalog, the sequences were then phase-folded.
The phase for each observation epoch, $\mathbf{t}$, was then replaced by the relative phase between the current and the previous epoch $\Delta \mathbf{t}$.
In particular, for the $j$-th measurement, $\Delta t_{j}=t_{j}-t_{j-1}$;
whereas $\Delta t_{0}=0$ is assumed for the first epoch.
The observed magnitudes in each light curve sequence $\mathbf{x}$ 
were normalized to have a mean of zero and a standard deviation of one, 
$\mathbf{x}\rightarrow (\mathbf{x}-\langle \mathbf{x}\rangle)/s$, 
where $\langle\mathbf{x}\rangle$ and $s$ are the mean and standard deviation of the sequence. 
The measurement errors were normalized by the same factor, 
$\boldsymbol{\sigma}\rightarrow \boldsymbol{\sigma}/s$. \newline

\begin{deluxetable*}{lclc}[t]
\tablecaption{Variable superclasses and their constituent subclasses.\label{tab:classnames}}
\tablecolumns{4}
\tablewidth{0pt}
\tablehead{
\colhead{Variable Type} & \colhead{Superclass Abbreviation} & \colhead{ASAS-SN Subclass\tablenotemark{a}} & \colhead{Number of Light Curve Sequences\tablenotemark{b}}}
\startdata
Cepheids                 & CEPH   &  CWA, CWB, DECP, DCEPS, RVA & 536 \\
Delta Scuti Variables    & DSCT   &  DSCT, HADS                 & 720 \\
Eclipsing Binaries       & ECL   &  EA, EB, EW                 & 25,713 \\
Mira Variables           & M   &  M                          & 1,020 \\
Rotational Variables     & ROT   &  ROT                        & 805 \\
RR Lyrae of Type A and B & RRAB  &  RRAB                      &  7,809 \\
RR Lyrae of Type C and D & RRCD   &  RRC, RRD                  &  3,014 \\
Semi-Regular Variables   & SR &  SR                        &  8,156 \\
\enddata
\tablenotetext{a}{Readers are referred to \citet{Jayasinghe+18b} and the ASAS-SN 
Variable Stars online database for the detailed descriptions of the subclasses.}
\tablenotetext{b}{Number of light curve sequences after data pre-processing. 
}
\end{deluxetable*}

\newpage

\subsection{Network Architecture}
A schematic diagram of the neural network in this work is shown in
Figure~\ref{fig:NN_schematic}.
It comprises of two sub-networks: \newline
   {\bf Autoencoder network:} \newline
        Feature extraction is performed using the autoencoding RNN in \citet{Naul+18}.
        The encoder and decoder each consist of two layers of
        Gated Recurrent Units (GRUs).
        A dropout layer\footnote{A dropout layer randomly ignores a fraction of units
        and their connections in a layer during training. 
        It improves robustness by preventing the network from overfitting with 
        sets of co-dependent weights. 
        The reduction in network size also reduces training time.}
        is included between the two recurrent layers to avoid overfitting \citep{Srivastava14}.
        Given a batch of $N$ normalized, phase-folded input light curve sequences 
        $(\Delta\mathbf{t}_{i},\mathbf{x}_{i},\boldsymbol{\sigma}_{i})$,
        where $i$ denotes the $i$-th light curve,
        the encoder converts them into reduced-dimension \emph{embedding vectors}, 
        $\mathbf{z}_{\textrm{e},i}\in\mathbb{R}^{m}$.
        The dimension of the embedding vector, $m$, is a user-defined parameter called the 
        \emph{embedding size}. 
        The embedding vectors are then used to reconstruct the input light curves
        $\hat{\mathbf{x}}_{i}$ by the decoder.
        Following \citet{Naul+18}, the loss function to be minimized is 
        the weighted mean square error,
        \begin{equation}
            L_\textrm{AE}=\frac{1}{n\;N}\sum_{i=1}^{N}\left(\mathbf{x}_{i}-\hat{\mathbf{x}}_{i}\right)^{2}\cdot\mathbf{w}_{i},
        \end{equation}
        where $\mathbf{w}_{i}=1/\boldsymbol{\sigma}_{i}$ is the sample weight for each epoch, 
        and the $\left(\cdot\right)^2$ is an element-wise operation.
        This loss function is advantageous over the conventional mean square error in that 
        it explicitly includes measurement errors from irregularly sampled observations.
        Through minimization of the deviation between the input and reconstructed light 
        curves, quantified by $L_{AE}$, the encoder is directed to produce embedding  
        vectors that contain representative information of the original light curves. \\

\begin{figure*}
  \centering
  \includegraphics[width=\columnwidth]{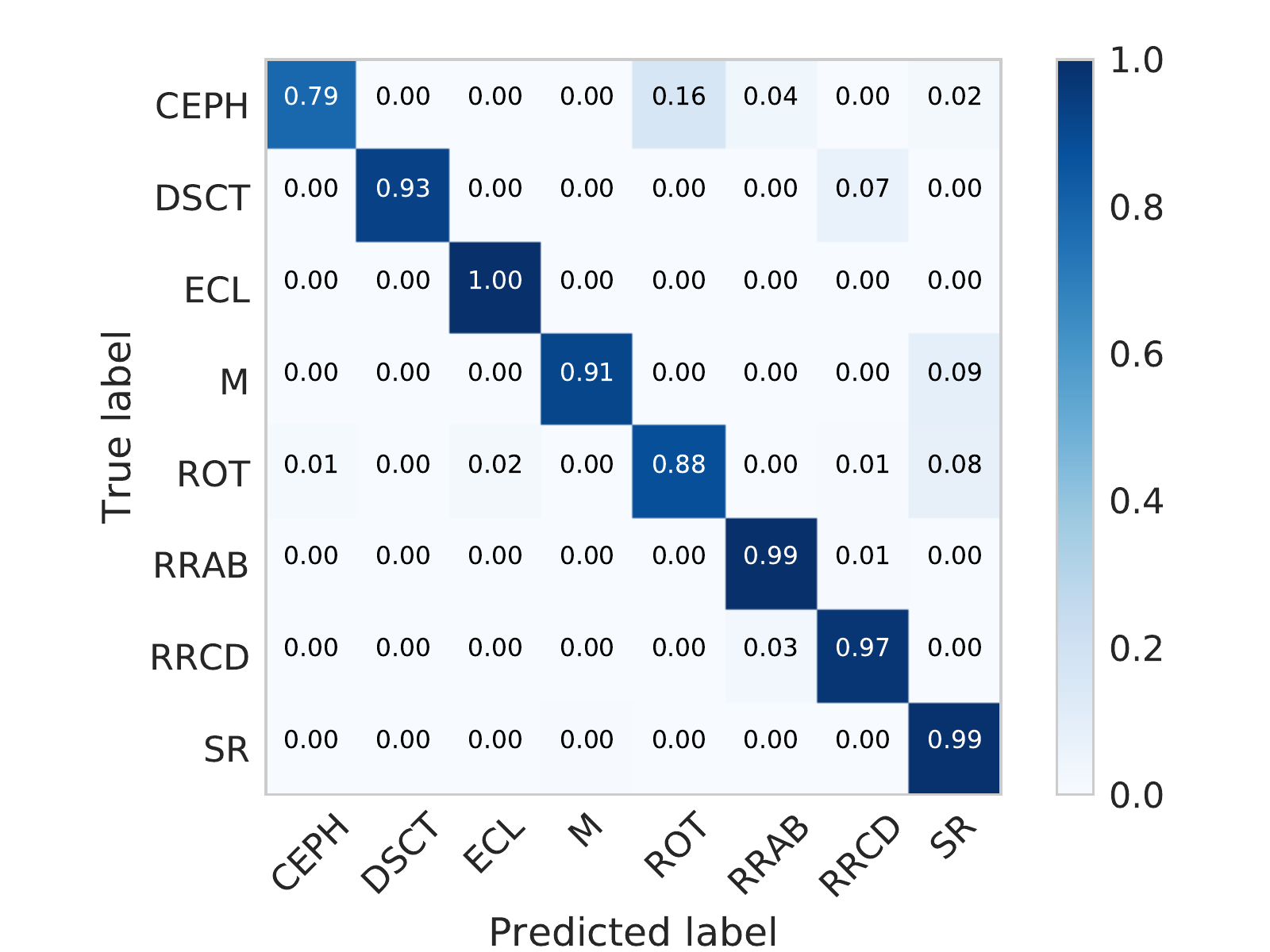}
  \includegraphics[width=\columnwidth]{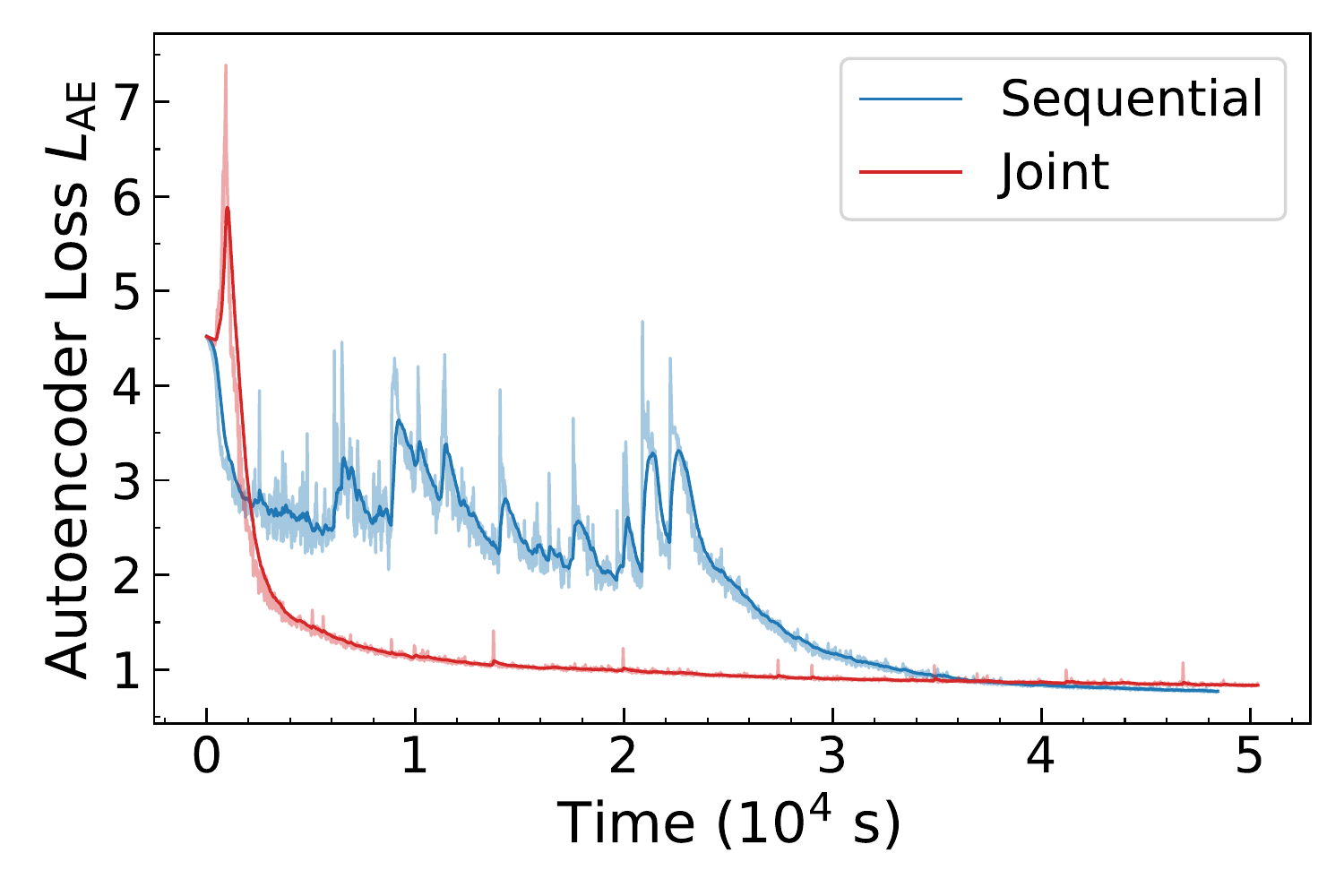}
  \caption{\emph{Left:} The normalized confusion matrix from the jointly trained autoencoder-estimation network.
  \emph{Right:} Time evolution of the autoencoder loss $L_\textrm{AE}$ computed from 
  the validation dataset during network training. 
  The solid lines show the linearly smoothed trends. 
  The lines with lighter colors shows the actual unsmoothed variations.
  }
  \label{fig:cm_loss}
\end{figure*}

    \noindent {\bf Estimation network:} \newline
        Classification and novelty detection are simultaneously accomplished by the estimation network.
        The input feature vector to the estimation network is constructed by combining the autoencoder embedding
        $\mathbf{z}_{\textrm{e},i}$, two additional reconstruction error features $\mathbf{z}_{\textrm{r},i}$, 
        and three auxiliary features $\mathbf{z}_{\textrm{a},i}$.
        Following \citet{Zong+18}, the two reconstruction error features are 
        the Euclidean distance and the cosine similarity, 
        \begin{equation}
            \mathbf{z}_{\textrm{r},i}=\left[\frac{\left\|\mathbf{x}_{i}-\hat{\mathbf{x}}_{i}\right\|_{2}}{\left\|\mathbf{x}_{i}\right\|_{2}},
                       \frac{\mathbf{x}_{i}\cdot\hat{\mathbf{x}}_{i}}{\left\|\mathbf{x}_{i}\right\|_{2}\left\|\hat{\mathbf{x}}_{i}\right\|_{2}}\right],
        \end{equation}
        where $|\cdot|_{2}$ denotes the $L_{2}$ norm. 
        The mean and standard deviation of each light curve sequence are combined with 
        the variable's period to form the auxiliary features, 
        $\mathbf{z}_{\textrm{a},i}=\left(\langle\mathbf{x}_{i}\rangle,s_{i},\log_{10}(P_{i})\right)$. 
        The input feature vector to the estimation network takes the form
        $\mathbf{z}_{i}~=~(\mathbf{z}_{\textrm{e},i},\mathbf{z}_{\textrm{r},i},\mathbf{z}_{\textrm{a},i})$,
        with dimension, $l=m+5$.
        
        The estimation network connects the input feature vector to an output layer
        through a single densely connected hidden layer. 
        The dimension of the output layer, $K$, is a pre-specified number of variable classes,
        which also corresponds to the number of Gaussian mixture components.
        A dropout layer is added after the hidden layer to minimize overfitting. 
        The output layer ends with a \emph{softmax} activation function,
        producing a $K$-dimensional, normalized vector, $\boldsymbol{\gamma}_{i}$, whose elements
        can be interpreted as the probabilities of belonging to each variable class/Gaussian component. 
        
        The classification functionality is trained by minimizing the categorical cross-entropy loss,
        \begin{equation}
          L_\textrm{CE}=-\frac{1}{N}\sum_{i=1}^{N}\mathbf{y}_{i}\cdot\log\left(\boldsymbol{\gamma}_{i}\right),
        \end{equation}
        where $\mathbf{y}_{i}$ is the true label of the $i$-th light curve expressed as 
        a $K$-dimensional one-hot vector\footnote{A one-hot vector is an integer vector with all but one element set to zero, with the non-zero element having a value of unity at the location denoting its membership 
        among one of the $K$ classes.}, 
        and the $\log$ is an element-wise natural logarithm on $\boldsymbol{\gamma_{i}}$.
        To generate the classification label predictions, the softmax outputs, $\boldsymbol{\gamma_{i}}$,
        are converted into one-hot vectors, $\hat{\mathbf{y}}_{i}$.
        
        The feature vectors, $\mathbf{z}_{i}$, and the estimation network outputs, 
        $\boldsymbol{\gamma}_{i}$, are used to compute the GMM parameters using 
        Equation (5) of \citet{Zong+18}.
        Following their notation, 
         $\boldsymbol{\mu}_{k}$ is the mean location, $\boldsymbol{\Sigma}_{k}$ is the 
        covariance matrix, and $\boldsymbol{\phi}_{k}$ is the me
        an membership probability 
        of the $k$-th Gaussian component in the feature space.
        The sample energy of each light curve sequence can be computed by
        \begin{equation}
            E(\mathbf{z}_{i})=-\log\left(\sum_{k=1}^{K}{\phi}_k\;\xi_k(\mathbf{z}_i)\right),
            \label{eqn:sample_energy}
        \end{equation}
        where $\xi_k$ is the normalized probability density in the $k$-th Gaussian component, 
        \begin{equation}
            \xi_k (\mathbf{z}_i)=\frac{\text{exp}(-\frac{1}{2}(\mathbf{z}_{i}-\boldsymbol{\mu}_k)^T
            \mathbf{\boldsymbol{\Sigma}}_k^{-1}(\mathbf{z}_{i}-\boldsymbol{\mu}_k))}{\sqrt{(2\pi)^{l}\mathbf{\boldsymbol{|\Sigma}}_k|}},
        \end{equation}
        $\log$ is the natural logarithm, and $|\cdot|$ denotes the matrix determinant.
        During network training, the GMM parameters are determined and 
        the associated loss is minimized using 
        \begin{equation}
            L_{\textrm{GMM}}=\frac{1}{N}\sum_{i=1}^{N}E(\mathbf{z}_{i}).
        \end{equation}
        Optimizing the estimation network by minimizing $L_\textrm{GMM}$ is equivalent to 
        fitting the GMM parameters by maximizing the log-likelihood.
        After training, the GMM parameters should adequately describe the distribution of variable types in the feature space. 
        New forms of variability can then be detected as outliers far away from 
        the Gaussian components.
        
        The overall estimation network architecture follows \citet{Zong+18} closely,
        but there are two main differences. 
        First, the network is tasked with the additional problem of  classification. 
        Second, with the addition of the classification loss, the GMM is not susceptible to 
        the singularity problem. The covariance loss is therefore unnecessary and omitted.

\begin{deluxetable*}{llcc}[t]
\tablecaption{Classification accuracy and novelty detection performance for the fiducial network 
with an embedding size of 16.
Numbers in parentheses correspond to results from networks trained using the additional photometric features.
\label{tab:main_results}}
\tablecolumns{4}
\tablewidth{0pt}
\tablehead{
\colhead{} & \colhead{} & \colhead{Joint Training} & \colhead{Sequential Training}}
\startdata
\multirow{2}{*}{Classification Accuracy} & Estimation Network &  98.8\% (99.1\%) & 96.7\% (96.6\%) \\
 & Random Forest      &  99.2\% (99.4\%) & 99.2\% (99.2\%) \\ 
\hline
\multirow{3}{*}{\shortstack[l]{Novelty Detection Scores  \\ 95th Percentile Cutoff}} & Precision  & 0.885 (0.898) & 0.875 (0.896)  \\
& Recall             & 0.815 (0.957) & 0.778 (0.933)  \\
& $F_{1}$ score      & 0.848 (0.927) & 0.824 (0.914) \\
\hline
\multirow{3}{*}{\shortstack[l]{Novelty Detection Scores  \\ 80th Percentile Cutoff}} & Precision  & 0.686 (0.700) & 0.683 (0.696)  \\
& Recall             & 0.941 (0.991) & 0.935 (0.986)  \\
& $F_{1}$ score      & 0.793 (0.821) & 0.789 (0.816) \\
\enddata
\end{deluxetable*} 

\subsection{Training Strategies}
The dual-network architecture was implemented using the Keras python package \citep{keras}
with a Tensorflow backend \citep{tf15}.
It was built on the GitHub implementation provided by \citet{Naul+18}.
The source code and a subset of the \textsc{ASAS-SN} data are available at 
\href{https://github.com/bthtsang/DeepClassifierNoveltyDetection}
{https://github.com/bthtsang/Deep\\ClassifierNoveltyDetection}.

Weights and biases in both networks are initialized with the \textsc{glorot\_uniform} initializer \citep{Glorot10}, 
and the loss function is minimized using the \textsc{Adam} optimizer \citep{KB14}.
We chose the same eight variable superclasses as used in \citet[Figure 29]{Jayasinghe+18b}
so that direct comparisons can be made between classification performance.
The variable subclasses that make up each superclass are summarized in 
Table \ref{tab:classnames}.

To demonstrate the versatility of the dual-network, 
two training approaches have been carried out,
namely, the \emph{joint} and \emph{sequential} training. 
In joint training, both the autoencoder and estimation network are optimized simultaneously
by minimizing the total loss,
\begin{equation}
  L_\textrm{tot}=L_\textrm{AE}+\left(\lambda L_\textrm{GMM}+L_\textrm{CE}\right),
  \label{eqn:L_tot}
\end{equation}
where the pre-factor $\lambda$ controls the relative importance of the GMM component. 
Different values of $\lambda$ have been tested and a fiducial value 
of $10^{-3}$ works well for the current application. 
For sequential training, the autoencoder is first trained utilizing only the $L_\textrm{AE}$ loss,
i.e. the exact training approach used in \citet{Naul+18}. 
The estimation network is trained afterwards, minimizing the parenthesized 
loss terms of Equation~(\ref{eqn:L_tot}). 
The motivation for including the sequentially trained models is twofold:
it produces an independently trained autoencoder with which we can benchmark our classification accuracy, 
it also provides an opportunity to assess the possibility of attaching novelty detection 
functionality on pre-trained feature extraction approaches.

An 80/20 split is used to divide the light curve sequences into \emph{training}
and \emph{validation} datasets.
To preserve the percentages of sequences in each variable class, 
the partition is performed using the \href{https://scikit-learn.org/stable/modules/generated/sklearn.model\_selection.StratifiedKFold.html}
{\textsc{StratifiedKFold}} function of the python \textsc{sklearn} package.
The validation dataset is withheld from the network during the entire training stage,
and is used to determine the classification accuracy of the networks. 

The following parameters are used in both training approaches.
A total of 96 GRUs were used in both layers of the RNN autoencoder network.
Training is done with a constant batch size of 2000 and learning rate of $2\times~10^{-4}$ for 2000 epochs. 
Dropout rates are fixed at 0.25 and 0.5 for the autoencoder and estimation network, respectively. 
We varied the embedding sizes between 8, 16, 32, and 64.
The size of the hidden layer in the estimation network is fixed at the embedding size.
As a benchmark, a separate grid of RF classifiers are trained using the 
encoder-generated features after network training. 
We adopted a grid of \texttt{n\_{estimators}} $\in\{50,100,250\}$, 
\texttt{criterion} $\in\{\textrm{gini},\textrm{entropy}\}$, 
\texttt{max\_features} $\in\{0.05,0.1,0.2,0.3\}$, 
and \texttt{min\_samples\_leaf} $\in\{1,2,3\}$ with the 
\textsc{SKLEARN} \href{https://scikit-learn.org/stable/modules/generated/sklearn.ensemble.RandomForestClassifier.html}{\textsc{RandomForestClassifier}} implementation.

\section{Results and discussions}
\label{sec:results}

\subsection{Classification Accuracy}
\label{sec:class_acc}
As embedding sizes of 8, 32 and 64 only offer marginally different performance in both classification and novelty detection,
in this Letter, we will focus on the results from the fiducial models 
with an embedding size of 16.
The normalized confusion matrix from the joint network training 
is shown on the left panel of Figure \ref{fig:cm_loss}.
The overall classification accuracy of the validation data is 98.8\%.
In particular, the classification performance on classes with copious amounts of light curves, 
namely ECL, RR Lyrae, and SR, is near-perfect.
With sequential training, the classification accuracy falls slightly to 96.7\%,
primarily due to misclassifications in the less populated superclasses. 

As a comparison, the best-performing RF classifier
gives an accuracy of 99.2\%, regardless of whether the network is trained jointly or sequentially.
The higher accuracy of RF is expected given its complexity relative to the basic estimation network.
In fact, the best-performing RF classifiers in the grid typically contain 
$\sim10^{2}$ decision trees, each consists of $\sim10^{2}$ nodes. 
A RF classifier therefore contains $\sim10^{4}$ trainable parameters, 
whereas the densely-connected layers of the estimation network have $\sim200-5000$, depending on the embedding size.
The classification accuracy and novelty detection performance scores for both
training approaches are summarized in Table \ref{tab:main_results}.

The right panel of Figure \ref{fig:cm_loss} shows the validation autoencoder loss 
over time during network training.
The joint training approach offers more efficient training of the autoencoder,
completing most of its learning early on after $10^{4}$\,s, 
or just $\sim$200 epochs.
Even though sequential training reaches slightly lower autoencoder loss
by the end of the training, classification accuracy is lower nevertheless. 
It suggests that by simultaneously optimizing both networks, 
the autoencoder is able to better retain class-specific information for a more effective
feature extraction. 
Across all embedding sizes (8, 16, 32, and 64), the joint training appeared to reduce stochastic fluctuations in the autoencoder loss, allowing superior, consistent training. 

\begin{figure*}
  \centering
  \includegraphics[width=\columnwidth]{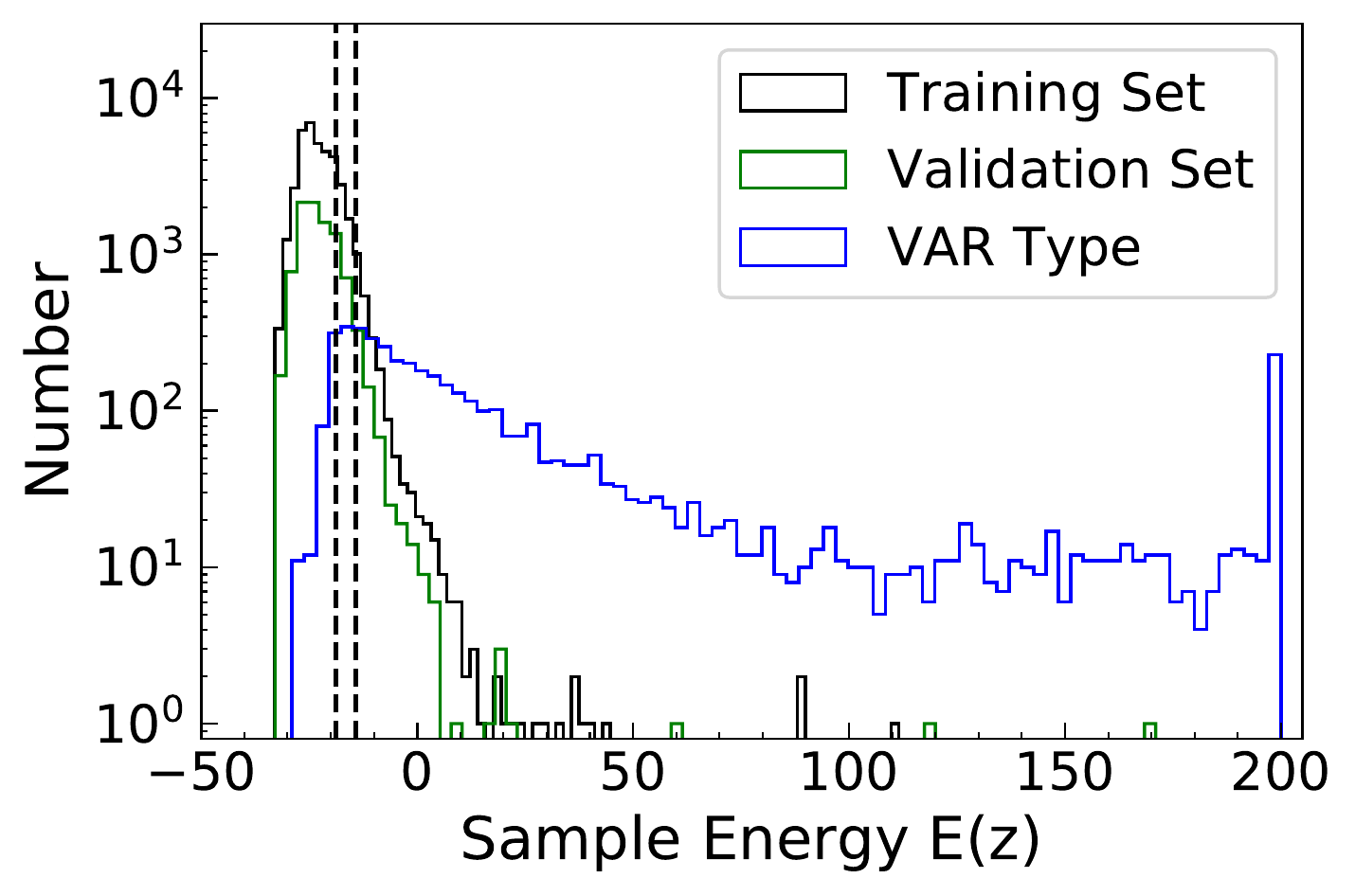}
  \includegraphics[width=\columnwidth]{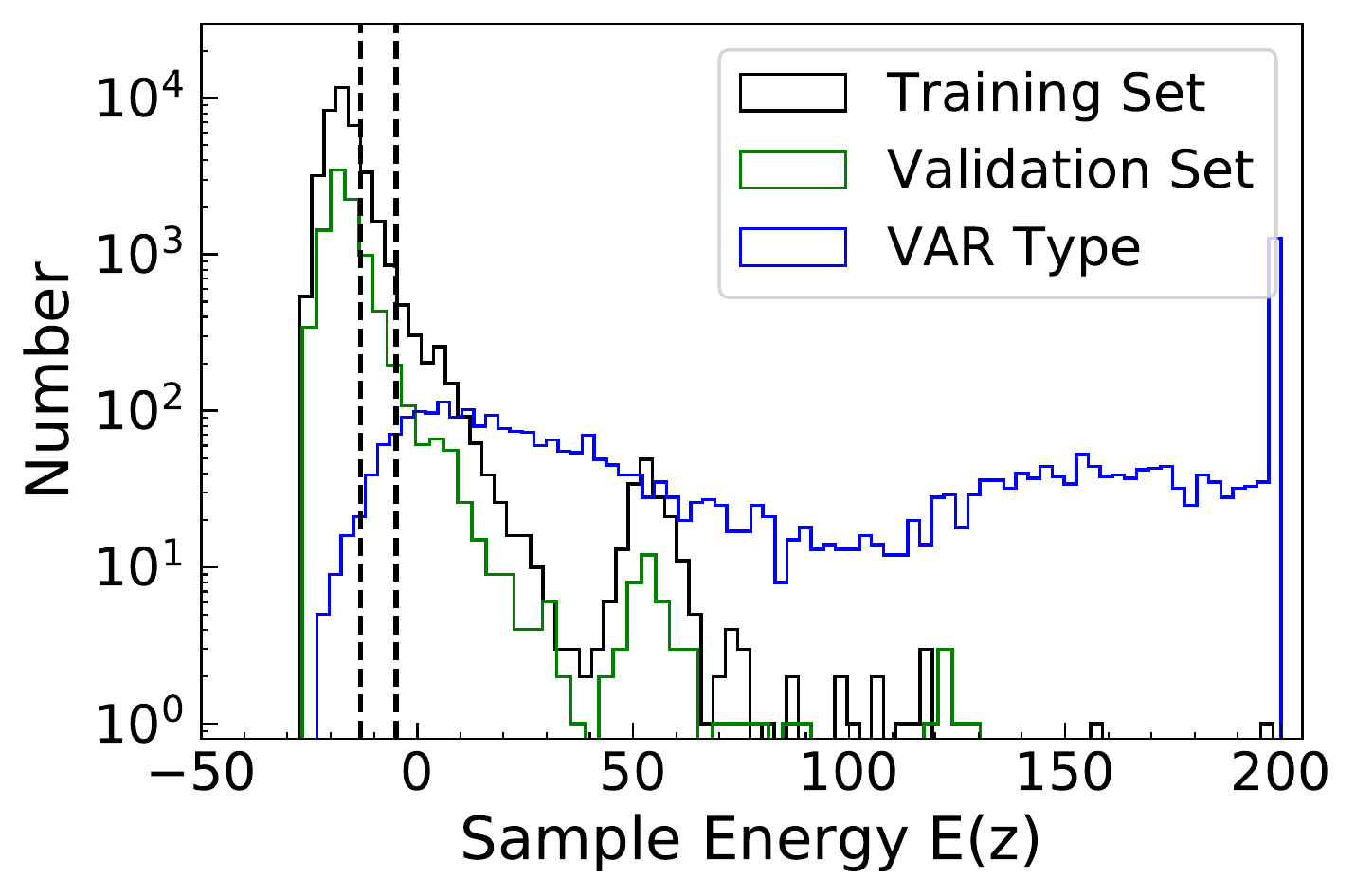}
  \caption{The energy histograms of the jointly trained network, without (left) and with (right) the inclusion of photometric features. An energy cap of 200 is imposed when creating the histograms for better visualization. 
  The vertical dashed lines denote the 80th and 95th percentiles of the training set. 
  The secondary peaks at $E\sim50$ in the case with included photometric features are the result of  
  missing magnitudes/colors, where a filling value of 99 is used.}
  \label{fig:energy_histogram}
\end{figure*}

\subsection{Novelty Detection Performance}
\label{sec:noveltydetection}

After the network training, the GMM parameters are computed and fixed using the entire 
training set.
The \textsc{VAR} light curve sequences are mixed with the validation dataset to form the 
test dataset for novelty detection, which contains objects both from the superclasses and of the
\textsc{VAR} type.
Light curve sequences from the test dataset were then passed into the encoder
to generate the embedding vectors. 
Together with the reconstruction error and auxiliary features, 
the sample energy of individual sequences can then be computed using Equation  (\ref{eqn:sample_energy}).
Light curves with sample energies above a pre-selected percentile cutoff in the training set, 
e.g. at 80\% or 95\%, are flagged as \emph{novel} samples.

The left panel of Figure \ref{fig:energy_histogram} shows the energy histogram of 
the light curve sequences in the training/validation dataset and of the \textsc{VAR} type.
Although there is a slight amount of overlap below $E\sim 0$, 
the majority of the \textsc{VAR} objects have higher energy and are visually distinct. 
Using a 95th percentile cutoff, the precision, recall, and $F_{1}$ score are 0.885, 0.815, 
and 0.848, respectively.
With sequential training, the performance scores are slight lower, 
at 0.875, 0.778, and 0.824 respectively. 
The reduction in recall implies that there are more \textsc{VAR} samples mixed
into samples with low energies and went undetected, suggesting 
that the GMM components are not as well-fitted. 
The performance scores are also summarized in Table~\ref{tab:main_results}.
Despite the lower accuracy and novelty detection scores, 
the performance of sequential training is satisfactory. 
It suggests that the estimation network can be readily integrated with existing 
classification pipelines with pre-computed features, e.g. from Fourier analysis, 
to deliver additional novelty detection functionality.

\subsection{Inclusion of Photometric Information}
\label{sec:photometric}
The majority of mis-classification in the confusion matrix (Figure \ref{fig:cm_loss}) appears among classes 
CEPH, M, and ROT, whose light curves are scarce and intrinsically similar to one another.
Though all classes benefited from explicitly including the auxiliary features 
(mean magnitude, standard deviation, variable period),
the three classes above require them for satisfactory classification accuracy.
Since the estimation network is responsible for classification \emph{and} novelty detection,
it is expected that improved classification accuracy will allow GMM components to be better optimized for enhanced novelty detection.

Photometric quantities from external catalogs are appended as additional features 
in an attempt to refine classification accuracies and thus improve novelty detection.
The photometric values were retrieved from the ASAS-SN Variable Stars Database
and concatenated as part of the input features to the estimation network. 
As our current intention is to examine the feasibility of incorporating 
additional photometric features, only three catalog quantities are selected, namely,
the Gaia DR2 $M_\textrm{G}$ magnitude, the $G_\textrm{BP}-G_\textrm{RP}$ color, 
and the Wesenheit Gaia $G_\textrm{RP}$ band magnitude $W_\textrm{RP}$.
The three confused classes are distinctly separated in this color-magnitude space
\citep[Figure 22]{Jayasinghe+18b}.
All parameters of the network were held fixed while training with photometric features so a direct comparison can be made. 

With the addition of the photometric information, the normalized accuracy in the 
CEPH, M, and ROT classes are boosted up to 84\%, 93\%, 93\% respectively.
The overall classification accuracy increases up to 99.1\%.
Most importantly, the novelty detection performance improved to 
reach an $F_{1}$ score of 0.93.
This exercise demonstrates that embedded light curve features can be trivially integrated 
with photometric features for better multi-task performance. 

\section{Conclusions}
\label{sec:conc}
We present a dual-network architecture that allows simultaneous training for
feature extraction, classification, and novel sample detection
of variable star light curves. 
We have combined the recurrent neural network autoencoder for time series data 
proposed by \citet{Naul+18} with the Gaussian mixture anomaly detection network 
proposed by \citet{Zong+18}. 
Applied to light curves from the \textsc{ASAS-SN} variable star database, 
the networks achieve a classification accuracy of $\sim$99\% and are able to detect
previously unseen types of variability with a precision, recall, and $F_{1}$ score
of $\geq 0.8-0.9$.

Joint training of the autoencoder and the classification/novelty detection
network is found to be mutually beneficial, 
resulting in more efficient autoencoder training and better overall performance. 
When trained on pre-extracted features, the network nevertheless produces
satisfactory results. It suggests that the Gaussian mixture-based network can be readily
integrated with existing classification pipelines for the added functionality of 
novelty detection.
Photometric features from external catalogs are found to be readily compatible with 
light curve features to deliver enhanced results. 

This work demonstrates the flexibility and extensibility of unsupervised
feature extraction of time series data for and beyond variable classification.
The dual-network architecture also highlights the fidelity of deep neural networks
in accomplishing multiple important tasks.

\section*{Acknowledgements}
This research project has benefited from interactions with Tom Prince, Thomas Kupfer, 
and Jan van Roestel.
The authors also thank Milos Milosavljevic and Tharindu Jayasinghe 
for valuable discussions and exchanges. 
We acknowledge support from the Center for Scientific Computing from the CNSI, 
MRL: an NSF MRSEC (DMR-1720256) and NSF CNS-1725797.
This research is funded by the Gordon and Betty Moore Foundation
through Grant GBMF5076 and supported in part by the National Science 
Foundation under Grant PHY-1748958 and by the NASA ATP grant ATP-80NSSC18K0560. 

\bibliographystyle{aasjournal}
\bibliography{letter}

\end{document}